\documentclass[preprints,editorial,accept,moreauthors]{mdpi} 
\usepackage{amsmath,amssymb,amsfonts,amsthm}

\firstpage{1} 
\pubvolume{6}
\issuenum{1}
\articlenumber{19}
\pubyear{2020}
\copyrightyear{2020}

\Title{Editorial for the Special Issue ``Progress in Group Field Theory and Related Quantum Gravity Formalisms''}

\newcommand{\ben}{\begin{equation}}
\newcommand{\een}{\end{equation}}

\newcommand{\bena}{\begin{eqnarray}}
\newcommand{\eena}{\end{eqnarray}}

\Author{Sylvain Carrozza $^{1}$\orcidA{}, Steffen Gielen $^{2}$\orcidB{}, Daniele Oriti $^{3}$\orcidC{}}

\AuthorNames{Sylvain Carrozza, Steffen Gielen, Daniele Oriti}

\address[3]{%
$^{1}$ \quad Perimeter Institute for Theoretical Physics, 31 Caroline St. N., Waterloo, ON N2L 2Y5, Canada; scarrozza@perimeterinstitute.ca
\\%
$^{2}$ \quad School of Mathematics and Statistics, University of Sheffield, Hicks Building, Hounsfield Road, Sheffield S3 7RH, UK; s.c.gielen@sheffield.ac.uk
\\%
$^{3}$ \quad Arnold-Sommerfeld-Center for Theoretical Physics, Ludwig-Maximilians-Universit\"at M\"unchen, Theresienstra\ss e 37, 80333 M\"unchen, Germany; daniele.oriti@physik.lmu.de}

\abstract{This Editorial introduces the Special Issue ``Progress in Group Field Theory and Related Quantum Gravity Formalisms'' which includes a number of research and review articles covering results in the group field theory (GFT) formalism for quantum gravity and in various neighbouring areas of quantum gravity research. We give a brief overview of the basic ideas of the GFT formalism, list some of its connections to other fields, and then summarise all contributions to the Special Issue.}


\begin{document}

\section{The group field theory formalism for quantum gravity}

Group field theory (GFT) sits at the intersection of various formalisms within the wider field of quantum gravity \cite{GFTreview1,GFTreview2,GFTreview3}. The basic idea behind GFT is to extend the framework of random matrix and tensor models, where a sum over triangulations is generated as the perturbative expansion of a theory of matrices or tensors, by including additional group-theoretic data to be interpreted as the discrete parallel transports of a connection formulation for gravity. These are the same variables that are fundamental to the definition of loop quantum gravity and spin foam models. GFT are thus a proposal for formulating the dynamics of quantum states built out of the kinematical data of loop quantum gravity, and thus for completing and extending the loop quantisation programme.

A straightforward example of a GFT that illustrates these aspects is the Boulatov model \cite{Boulatov} for three-dimensional Riemannian quantum gravity. This model is defined by the action
\bena
S_{{\rm Boul}}[\varphi]&=&\frac{1}{2}\int {\rm d}^3 g\,\varphi^2(g_1,g_2,g_3)\nonumber
\\&&-\frac{\lambda}{4!}\int {\rm d}^6 g\,\varphi(g_1,g_2,g_3)\varphi(g_1,g_4,g_5)\varphi(g_2,g_5,g_6)\varphi(g_3,g_6,g_4)\,,
\eena
where the GFT field $\varphi$ is a real-valued function on three copies of ${\rm SU}(2)$ with an additional permutation symmetry,
\ben
\varphi: {\rm SU}(2)^3 \rightarrow \mathbb{R}\,,\quad \varphi(g_1,g_2,g_3)=\varphi(g_2,g_3,g_1)=\varphi(g_3,g_1,g_2)\,,
\een
and ``gauge invariance'' under the diagonal left action of the group on all the arguments of the field,
\ben
 \varphi(g_1,g_2,g_3)= \varphi(hg_1,hg_2,hg_3)\quad \forall h\in{\rm SU}(2)\,.
\een
The action consists of a quadratic ``kinetic'' term, with trivial propagator, and an interaction term with a somewhat unusual (``nonlocal'') pairing of arguments. In fact, the group nature of the domain of the dynamical fields and such non-local pairing of arguments in the interactions (shared with matrix and tensor models) can be understood as defining properties of the formalism, the other ingredients (e.g., symmetries, choice of group, kinetic and interaction terms) being  a specification of models within the general framework. If one now considers the perturbative expansion of the partition function
\ben
Z_{{\rm Boul}} = \int \mathcal{D}\varphi\;e^{-S_{{\rm Boul}}[\varphi]}
\een
in powers of the coupling $\lambda$, due to this peculiar structure of the interaction term, the Feynman graphs arising in such an expansion are dual to three-dimensional simplicial complexes, i.e., discrete combinatorial spacetimes. Concretely, one finds
\ben
Z_{{\rm Boul}}=\sum_{\Gamma} \lambda^{V(\Gamma)}\sum_{\{j_f\}\in{\rm Irrep}}\prod_{f\in \Gamma} (2j_f+1)\prod_{v\in \Gamma}\left\{ \begin{matrix} j_{v_1} & j_{v_2} & j_{v_3} \cr j_{v_4} & j_{v_5} & j_{v_6} \end{matrix}\right\}\,,
\label{pertexp}
\een
which is a sum over graphs $\Gamma$ and, for each $\Gamma$, over assignments of irreducible representations $j_f$ of ${\rm SU(2)}$ to each face of $\Gamma$. For each such assignment of $j_f$ one finds an amplitude which is a product over `face amplitudes' $(2j_f+1)$ and `vertex amplitudes' given by a Wigner $6j$-symbol for the six faces that meet at a vertex (the number six arising from the six group elements integrated over in the interaction).

Each graph $\Gamma$ is dual to an oriented 3d simplicial complexes $C$, where each vertex $v\in\Gamma$ is dual to a tetrahedron $T\in C$, and each face $f\in\Gamma$ dual to a link $l\in C$. The interesting observation is now that the amplitude appearing in the expansion (\ref{pertexp}) is nothing but the {\em Ponzano--Regge state sum} \cite{PonzanoRegge} of the triangulated manifold $C$, multiplied with an overall weight $\lambda^{V(\Gamma)}\equiv \lambda^{N_T(C)}$ depending on the number $N_T(C)$ of tetrahedra in $C$. The Ponzano--Regge state sum defines a discrete path integral for three-dimensional quantum gravity on a given triangulation $C$, written in a basis which is the analogue of spherical harmonics; see, e.g., \cite{PonzanoRegge2} for details and a discussion of how to rigorously define such a state sum. In these variables, one obtains what is known as a spin foam model in the loop quantum gravity literature, i.e., a covariant definition of the quantum dynamics of spin networks. The same Feynman amplitudes can also be expressed directly in group variables, where they take the form of a lattice gauge theory for 3$d$ BF theory (equivalent to pure 3$d$ gravity with no cosmological constant). An expression in which the same amplitudes coincide with the discrete path integral for 3$d$ quantum gravity in triad and connection variables can also be given. 

In summary, the perturbative expansion of the Boulatov model generates a sum over discrete (simplicial) spacetimes with a discrete quantum gravity path integral assigned to each spacetime, augmented by a sum over discrete topologies:
\ben
Z_{{\rm Boul}} = \sum_C \lambda^{N_T(C)} Z_{{\rm PR}}(C)\,.
\een
The Ponzano--Regge state sum defines a topological field theory which is triangulation independent for fixed topology, so that the sum over triangulations of the same $3d$ manifold merely leads to repeated factors of the same state sum appearing in this expansion. However, in models which are not topological, summing over {\em all} simplicial complexes would restore discretisation independence.

This correspondence between perturbative expansions of appropriate GFT models and discrete path integrals for quantum gravity, as well as the connection to spin foam models, extend to other cases, in particular candidate models for quantum gravity in four spacetime dimensions. Spin foam models of immediate interest for  loop quantum gravity \cite{SF}, for which a more detailed understanding in terms of simplicial geometry is available, can be obtained from the expansion of a GFT for a field with four arguments valued in the Lorentz group or SU(2), with additional geometricity conditions imposed on the kinetic or interaction kernels and combinatorially nonlocal interactions of $\varphi^5$ type \cite{GFTSF}. 

In what we have presented so far, the GFT approach appears to give simply a reformulation of known expressions for spin foam models and other discrete quantum gravity path integrals, which can be obtained by other means. However, being able to define them in terms of a quantum field theory --- albeit an unusual one in that the field $\varphi$ does not live on spacetime but on an abstract group manifold --- provides further avenues to explore spin foam and other models, beyond studying the GFT perturbative expansion. 

In particular, one can study perturbative and non-perturbative renormalisation of GFTs, and look for theories that can be defined consistently at all scales, and hence become candidates for a fundamental theory. Relying on results and methods developed in the context of tensor models \cite{tensor} which share the same basic combinatorial structure as GFTs, a tentative but mathematically precise GFT renormalisation framework has been developed \cite{tensor_track, review_ren}. It has allowed to demonstrate the perturbative renormalisability --- that is, the consistency and predictivity --- of simple but non-trivial `tensorial' GFT actions, and has led to further investigations of the phase structure of GFTs at the non-perturbative level. While the precise physical interpretation of such abstract and background-independent fixed points remains to be elucidated, it is hoped that these technical advances will find suitable extensions to realistic four-dimensional GFT models of quantum gravity.

Regarding the perturbative expansion of GFT itself, tensorial GFT actions are of even broader interest because they admit a $1/N$ expansion. As in the widely studied case of matrix models for two-dimensional quantum gravity, the $1/N$ expansion allows to partially resum the perturbative expansion, and thereby provides crucial control over the critical regime of the theory. As a result, the study of tensorial GFT models has seen a number of interesting developments in recent years \cite{tensor_track-III}.

The renormalisation analysis is also related to the search for a continuum limit in GFTs which can be pursued with quantum field theory methods, addressing the key open question of a continuum limit (or sum over discretisations) in spin foam models and loop quantum gravity. This continuum limit may be given by a non-perturbative phase (often suggested to be of condensate type) in which the GFT field acquires a nonvanishing expectation value, which would be where relevant continuum physics is found (in particular, the number of building blocks diverges). The possible condensate phase of GFTs has been studied with methods coming from condensed matter theory, and has been applied to the description of cosmology and black holes within GFT \cite{condensate_review, condensate_reviewII}. For example, the emergent cosmological dynamics of the universe, whose microscopic description is given by a GFT model, is extracted from the condensate hydrodynamics rephrased in terms of suitable geometric observables. These effective cosmological dynamics show both the correct classical limit at large volumes and a rather generic bouncing dynamics in place of the classical big bang singularity. Moreover, under further assumptions, they match the effective dynamics found in loop quantum cosmology.

Motivated by these applications to cosmology, formal investigations of the algebraic structure of GFTs have been initiated, aiming at a more refined account of GFT condensate states, and of the condensation mechanism itself. Even more ambitiously, this research direction lays the groundwork for a reformulation and extension of thermal physics to background-independent quantum gravity \cite{isha_daniele,isha_daniele2}.

This Special Issue consists of contributions related to the different avenues of research within the GFT programme and to neighbouring areas of interest. As we have made clear, loop quantum gravity, spin foam models, and more generally discrete and combinatorial approaches to quantum gravity are closely related to GFT and thus work in these fields has direct implications for GFT. Vice versa, results in the GFT formalism could be of both inspiration and direct application in other quantum gravity formalisms. Looking further afield, submissions from research fields with relevance to more specific aspects of GFT research were also encouraged; these included for instance fundamental cosmology, quantum information or condensed matter theory, but also mathematical and formal aspects.

\section{Contributions to the Special Issue}

The Special Issue consists of 14 published manuscripts; ten research articles and four review articles. The research articles (listed in chronological order of publication) cover the following topics:

\begin{itemize}

\item A number of symmetry-reduced models of loop quantum gravity (LQG) have indicated that the fine structure of the LQG quantum state space may naturally lead to deformations of the constraint algebra of general relativity at the semiclassical level. This can in turn be interpreted as a quantum deformation of general covariance, required by the existence of a new invariant length scale, the Planck scale. In {\em Rainbow-Like Black-Hole Metric from Loop Quantum Gravity} \cite{rainbow}, Iarley P. Lobo and Michele Ronco investigate spherically-symmetric black hole solutions predicted by effective models of LQG. They show that their quantum-deformed covariance leads to a modified dispersion relation for the total radial momentum, which they then analyse within the paradigm of rainbow gravity. 

\medskip

\item {\em Primordial Power Spectra from an Emergent Universe: Basic Results and Clarifications} \cite{powersp} by Killian Martineau and Aur\'elien Barrau discusses a non-standard scenario for the beginning of the universe, known as the emergent universe. In the emergent universe, the Big Bang (or big bounce) is replaced by a transition from a static to an expanding universe. The authors investigate features of the primordial power spectrum of tensor perturbations, or gravitational waves from the early universe. They study conditions that are required for a scale-invariant spectrum from an emergent universe scenario and show how features of the spectrum depend on the details of the scale factor evolution near the transition from static to expanding phase.

\medskip

\item One of the most ambitious hopes for quantum gravity is that it can teach us something about the initial state of the universe. {\em On the Geometry of No-Boundary Instantons in Loop Quantum Cosmology} \cite{instant} takes up one of the most prominent ideas of this type, Hawking's no-boundary proposal, and incorporates quantum corrections from loop quantum cosmology into it. Suddhasattwa Brahma and Dong-han Yeom study semiclassical instanton solutions to the LQC path integral. They find that, in contrast to calculations in pure semiclassical general relativity, these instantons have a characteristic infinite tail, and they tend to close off in a regular way as was one of the original ideas behind the no-boundary proposal.

\medskip

\item In {\em Equivalence of Models in Loop Quantum Cosmology and Group Field Theory} \cite{equiv}, Bekir Baytas, Martin Bojowald and Sean Crowe observe that the emergent GFT dynamics of homogeneous isotropic universes filled with a massless scalar, which form the basis of the application of GFT to cosmology, can be understood in terms of the algebraic structure of the Lie algebra $\mathfrak{su}(1,1)$. The same algebra structure is known to underlie the most studied models of loop quantum cosmology. The similarities seen between cosmological features of GFT and loop quantum cosmology are then explained in algebraic terms. Furthermore, this underlying algebraic structure suggests possible generalisations of GFT cosmology.

\medskip

\item In {\em Status of Background-Independent Coarse Graining in Tensor Models for Quantum Gravity} \cite{CoarseG}, Astrid Eichhorn, Tim Koslowski and Antonio D. Pereira explore applications of the functional renormalisation group to tensor models. They review recent efforts attempting to leverage non-perturbative methods to probe the existence of new large-$N$ limits in tensor models. Once rephrased in the appropriate renormalisation group language, in which the size of the tensor plays the role of abstract scale, the existence of such a scaling limit manifests itself by the presence of a non-trivial renormalisation group fixed point. The Wetterich equation then provides an elegant and powerful discovery tool, which allows to scan the theory space of tensor models within larger and larger truncations. From the point of view of quantum gravity, any new large-$N$ limit will translate into a new way of taking the continuum limit. Such investigations are therefore crucial for assessing the viability of tensor and GFT models of quantum gravity in dimension higher than two. 

\medskip

\item {\em Reconstruction of Mimetic Gravity in a Non-Singular Bouncing Universe from Quantum Gravity} \cite{mimetic} by Marco de Cesare deals with bouncing cosmologies such as have been found in the application of GFT to cosmology. Such bouncing cosmologies have also been seen in models of (limiting curvature) mimetic gravity, in which one modifies gravity by including a scalar field; therefore, the precise relation of mimetic gravity and the cosmological sector of quantum gravity has recently attracted interest. This paper presents a reconstruction procedure by which, starting from a given cosmological effective dynamics from quantum gravity, one can obtain a classical mimetic gravity action (given in terms of a particular function $f(\Box\phi)$) that reproduces this cosmological solution, in the isotropic and homogeneous sector. This might then be seen as a candidate for an effective field theory for quantum gravity approaches such as GFT. The effective field theory is then used to study anisotropies and inhomogeneities.

\medskip

\item Philipp A. H\"ohn's article {\em Switching Internal Times and a New Perspective on the `Wave Function of the Universe'} \cite{switch} discusses the fundamental question of how to extend the notion of general covariance from classical to quantum gravity. The central question is how to switch between descriptions given by different observers of what should be the same physics; in other words, between quantum reference frames. Such a relational definition of the quantum dynamics is commonly employed in quantum gravity, and, for example, in GFT cosmology, to define evolution of geometric quantities in a fully diffeomorphism invariant, thus physical, manner. The paper formulates a general method for relating reduced quantum theories (theories defined after a choice of reference system) to the perspective-neutral framework of Dirac quantisation, akin to the passage from a given coordinate system to generally covariant expressions in classical general relativity. This is then applied to simple models of quantum cosmology where it provides a new angle on the `wave function of the universe', which becomes a global, perspective-neutral state, encoding all descriptions of the universe relative to different choices of reference system.

\medskip

\item The study of cosmological perturbations is important in the application of quantum gravity models to the early universe, including, for example, in the context of GFT cosmology. {\em Dynamical Properties of the Mukhanov--Sasaki Hamiltonian in the Context of Adiabatic Vacua and the Lewis--Riesenfeld Invariant} \cite{MuSas} by Max Joseph Fahn, Kristina Giesel and Michael Kobler aims to define suitable initial quantum states for inflation in a near-de Sitter geometry using Hamiltonian methods. The dynamics of cosmological perturbations in an expanding universe can be written in the form of harmonic oscillators with time-dependent frequency. For finite-dimensional systems with such dynamics, an important role is played by the Lewis--Riesenfeld invariant, a constant of motion. One of the main aims of this paper is to extend the application of the Lewis--Riesenfeld invariant to the infinite-dimensional case of field theory. The states thus generated as candidates for an initial state for inflation are then compared to well-known initial states such as the Bunch--Davies vacuum.

\medskip

\item {\em Spin Foam Vertex Amplitudes on Quantum Computer---Preliminary Results} \cite{SFV} by Jakub Mielczarek outlines first steps of an ambitious project: the use of quantum algorithms to understand spin foam vertex amplitudes, one of the key ingredients in defining the dynamics of spin foam models (and hence indirectly, of GFT models). In this article, the focus is on a simple spin network (a complete graph of five vertices representing five tetrahedra forming the boundary of a 4-simplex) with all spins set equal to $\frac{1}{2}$. The paper discusses how to calculate absolute values of vertex amplitudes for this process, and the approach is tested by comparing the results obtained by existing quantum algorithms with known exact results.

\medskip

\item In {\em Thermal Quantum Spacetime} \cite{TQST}, Isha Kotecha discusses an extension of equilibrium statistical mechanics and thermodynamics to background-independent systems that is then applied to discrete quantum gravity approaches such as GFT. A generalised notion of Gibbs equilibrium is characterised in information-theoretic terms, where entropy plays a more fundamental role than energy. This then forms the basis for a framework of a statistical mechanics of discrete quantum gravity in the absence of standard notions of time and energy. Covariant GFT is shown to arise as an effective statistical field theory of generalised Gibbs states. The paper presents also a conceptual review of these and other results in this context and an extensive outlook of further work in this important direction.
\end{itemize}

The Special Issue also includes four review articles, namely

\begin{itemize}
\item In {\em Quantum Gravity on the Computer: Impressions of a Workshop} \cite{workshop}, Lisa Glaser and Sebastian Steinhaus summarise the outcome of the workshop they organised in March 2018 at NORDITA, in Stockholm. Spanning a rather wide array of distinct approaches (including loop quantum gravity and spin foams, as well as GFT), this article reviews recent and ongoing contributions of computational physics to open problems in discrete quantum gravity, such as those related to the challenging question of the restoration of the diffeomorphism symmetry in the continuum limit. The review concludes with an insightful roadmap, which among other targets advocates the creation of open data science infrastructures and online repositories dedicated to numerical investigations of quantum geometry.  

\medskip

\item Functional renormalisation group (FRG) techniques have recently been successfully applied to GFT models. {\em Progress in Solving the Nonperturbative Renormalization Group for Tensorial Group Field Theory} \cite{RGreview} by Vincent Lahoche and Dine Ousmane Samary gives an overview over three previous papers by these authors, in which the FRG is applied to Abelian GFT models based on gauge group $U(1)^d$, without a closure/gauge invariance constraint (such a constraint is usually imposed for the geometric interpretation of these models, as it introduces a gauge connection and turns GFT models into a quantisation of gauge theories or gauge-theoretic gravitational models). A quartic interaction term of melonic type is studied in these models. An effective vertex expansion method is introduced in order to solve the FRG and study the resulting renormalisation flow, in particular with the aim of identifying non-Gaussian fixed points; these fixed points may be associated to phase transitions that can be interpreted as describing the formation of a GFT condensate (see above). Ward--Takahashi identities provide additional constraints that have to be taken into account when finding approximate solutions to the flow equations.

\medskip

\item In recent years, the GFT formalism has permitted the emergence of a new approach to quantum cosmology, based on the general paradigm of condensation in GFT, as we discussed above. Thanks to the quantum field theory language underlying GFT, the idea that cosmological spacetime structures may be the result of the condensation of a large number of pre-geometric and quantum degrees of freedom has been concretely realised and thoroughly investigated in simple GFT models. In {\em Group Field Theory Condensate Cosmology: An Appetizer} \cite{appetizer}, Andreas G.~A.~Pithis and Mairi Sakellariadou provide a gentle and pedagogical introduction to this fast-developing area of research. After reviewing how isotropic and homogeneous cosmology can be recovered from a GFT condensate, they summarise recent efforts aiming at including anisotropies and cosmological perturbations, paving the way towards the derivation of observable consequences. 

\medskip

\item A number of recent developments in quantum gravity suggest that the Einstein equations might be best understood as a reflection of the entanglement structure of fundamental and yet-to-be-discovered quantum gravity degrees of freedom. In the context of the AdS/CFT correspondence, this idea is beautifully captured by the Ryu-Takayanagi formula, which relates the entanglement entropy of regions in the boundary CFT to the area of extremal surfaces in the bulk. In {\em Holographic Entanglement in Group Field Theory} \cite{holo}, Goffredo Chirco reviews the realisation of such ideas in the context of GFT, where candidate microscopic degrees of freedom are available. Relying on a general dictionary allowing to view GFT many-body states as tensor networks, a pedagogical introduction to the computation of R\'enyi entropies by means of the replica method is proposed. This allows the author to derive a GFT analogue of the Ryu-Takayanagi equation, which is fully compatible with the geometric interpretation of the GFT fundamental degrees of freedom: the area term entering the formula is consistently given by the expectation value of the corresponding GFT area operator.  
\end{itemize}

\funding{The work of SG is supported by the Royal Society under a Royal Society University Research Fellowship (UF160622) and a Research Grant for Research Fellows (RGF\textbackslash R1\textbackslash 180030). The work of DO is supported by the Deutsche Forschung Gemeinschaft (DFG).}

\acknowledgments{The guest editors would like to thank all the authors for their contributions and the reviewers for the constructive reports. Their work helped the editors to collect this Special Issue. SC acknowledges support from Perimeter Institute. Research at Perimeter Institute is supported in part by the Government of Canada through the Department of Innovation, Science and Economic Development Canada and by the Province of Ontario through the Ministry of Economic Development, Job Creation and Trade.}

\conflictsofinterest{The authors declare no conflict of interest. The funders had no role in the writing of the manuscript, or in the decision to publish the results.}

\reftitle{References}

\end{document}